\documentclass{webofc}
\usepackage{lipsum}
\usepackage{lineno}
\usepackage{placeins}
\usepackage[varg]{txfonts}   
\begin{document}
%
\title{Chasing the onset of QCD thermalization with ALICE}
%
%

\author{\firstname{Mario} \lastname{Ciacco}\inst{1}\fnsep\thanks{\email{mario.ciacco@cern.ch}} on behalf of the ALICE Collaboration
}

\institute{Dipartimento DISAT del Politecnico and Sezione INFN, Turin, Italy}

\abstract{%
  In heavy-ion and hadronic collisions, indications of thermalization are detected in the yields of produced hadrons: these observations call for a detailed study of the hadronization processes. Novel observables are required to discriminate between the different hadronization models that successfully describe the average yields of hadrons. The ALICE Collaboration reports on the studies of event-by-event fluctuations of the net-$\Xi^\pm$ number, and Pearson correlations of both the net-$\Xi^\pm$--net-kaon numbers and the antideuteron--antiproton numbers, in pp, p--Pb, and Pb--Pb collisions. The measured fluctuations are compared with different hadronization models to explore the compatibility with a charge-equilibration scenario. In all cases, the experimental data agree with Thermal-FIST canonical-statistical calculations, allowing for the extraction of the correlation length for strange and baryon quantum number conservation.
}
\maketitle

\section{Introduction}
The production of hadrons in heavy-ion collisions is described by statistical hadronization models (SHM), in which the system of produced particles is described as a thermalized ideal gas of light-flavored hadrons and resonances~\cite{Andronic17}. Over the past years, this approach has also been applied in smaller collision systems, such as pp and p--Pb, taking additionally into account effects of partial thermal equilibration of the more-massive strange quark~\cite{Vovchenko19}. The agreement between the experimental data and SHM models poses the question of the onset of the equilibration of such hadronic systems. This is especially relevant in the study of strangeness hadronization and (anti)nucleosynthesis in hadronic and heavy-ion collisions. A continuous evolution of the strange hadron to charged pion yield ratios on the one hand~\cite{AliceNature}, and of the (anti)nuclei to proton yield ratios on the other~\cite{AliceNuclei}, is observed going from small to large collision systems. These results are described both by canonical SHM models and by models based on the Lund string-fragmenation mechanism. Specifically, the string fragmentation model of Pythia~\cite{Pythia}, including both the interactions among strings through the rope hadronization mechanism~\cite{PythiaRope} and baryon enhancement effects generated by the formation of string junctions~\cite{PythiaCR}, provides an accurate description of strange-hadron average yields. Concerning (anti)nucleosynthesis, light (anti)nuclei yields can be alternatively described either by SHM models~\cite{VovchenkoNuclei} or by the nuclear coalescence model~\cite{KJSunCoalescence}, according to which bound states are formed when nucleons emitted at a freeze-out hypersurface are close in phase space.

Novel observables are needed to discriminate among different approaches. In this contribution, the ALICE Collaboration reports on the measurements of event-by-event fluctuations in the strangeness~\cite{PublicNote} and light-nuclei sectors~\cite{AntidFluct}. The studied observables are extracted from cumulants, defined as $\kappa_{i}=\langle (n - \langle n \rangle)^i\rangle$ for $i<3$, with $n$ being the event-by-event particle number and $\langle\rangle$ indicating the average over events. First and second order cumulants correspond to the average and variance, while the mixed cumulant $\kappa_{11}$ is the covariance, thus allowing us to extract Pearson correlations, $\rho$, between different particle species~\cite{PublicNote}.

\begin{figure}
\centering
\includegraphics[width=0.48\textwidth,]{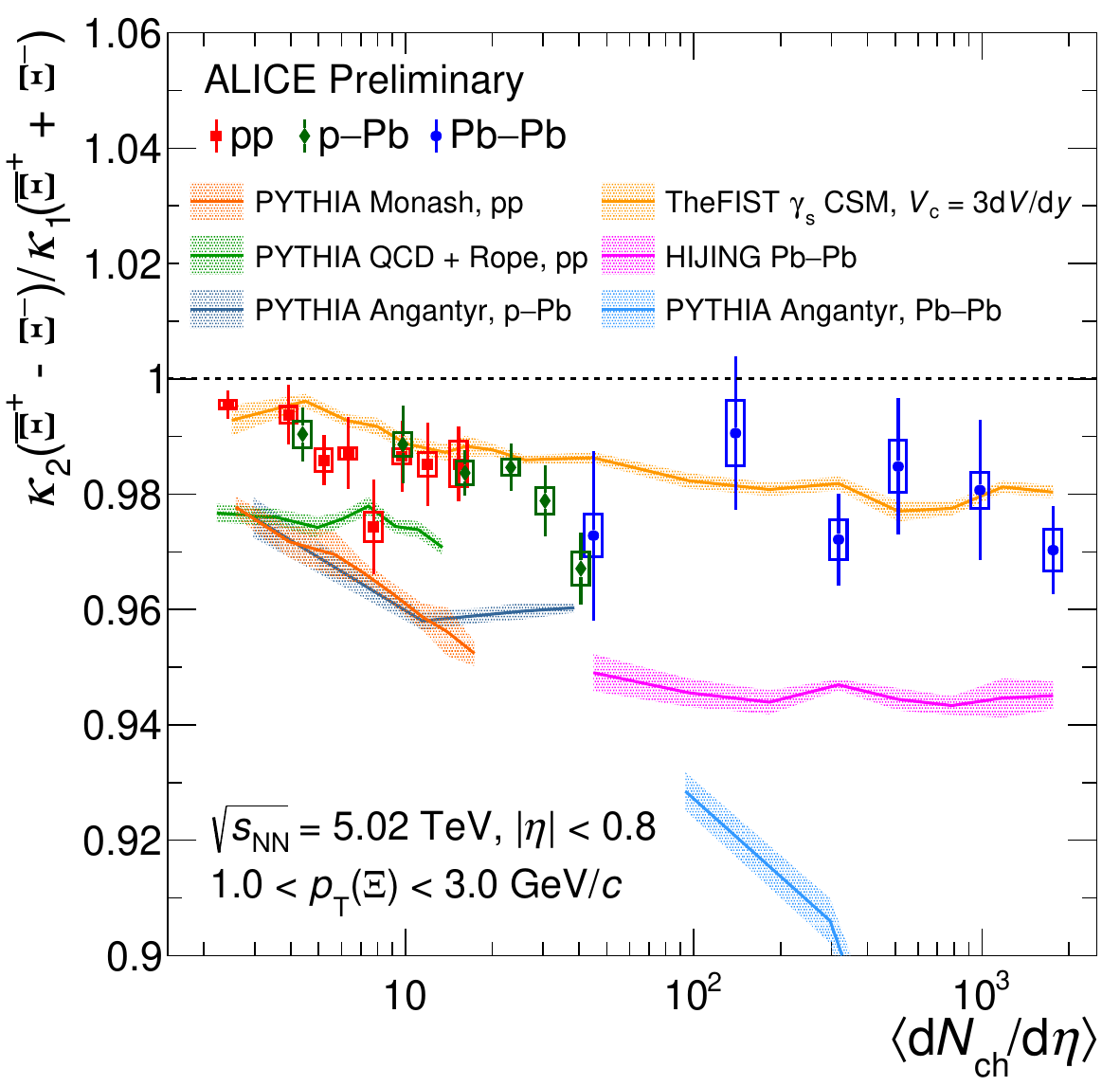}
\includegraphics[width=0.48\textwidth]{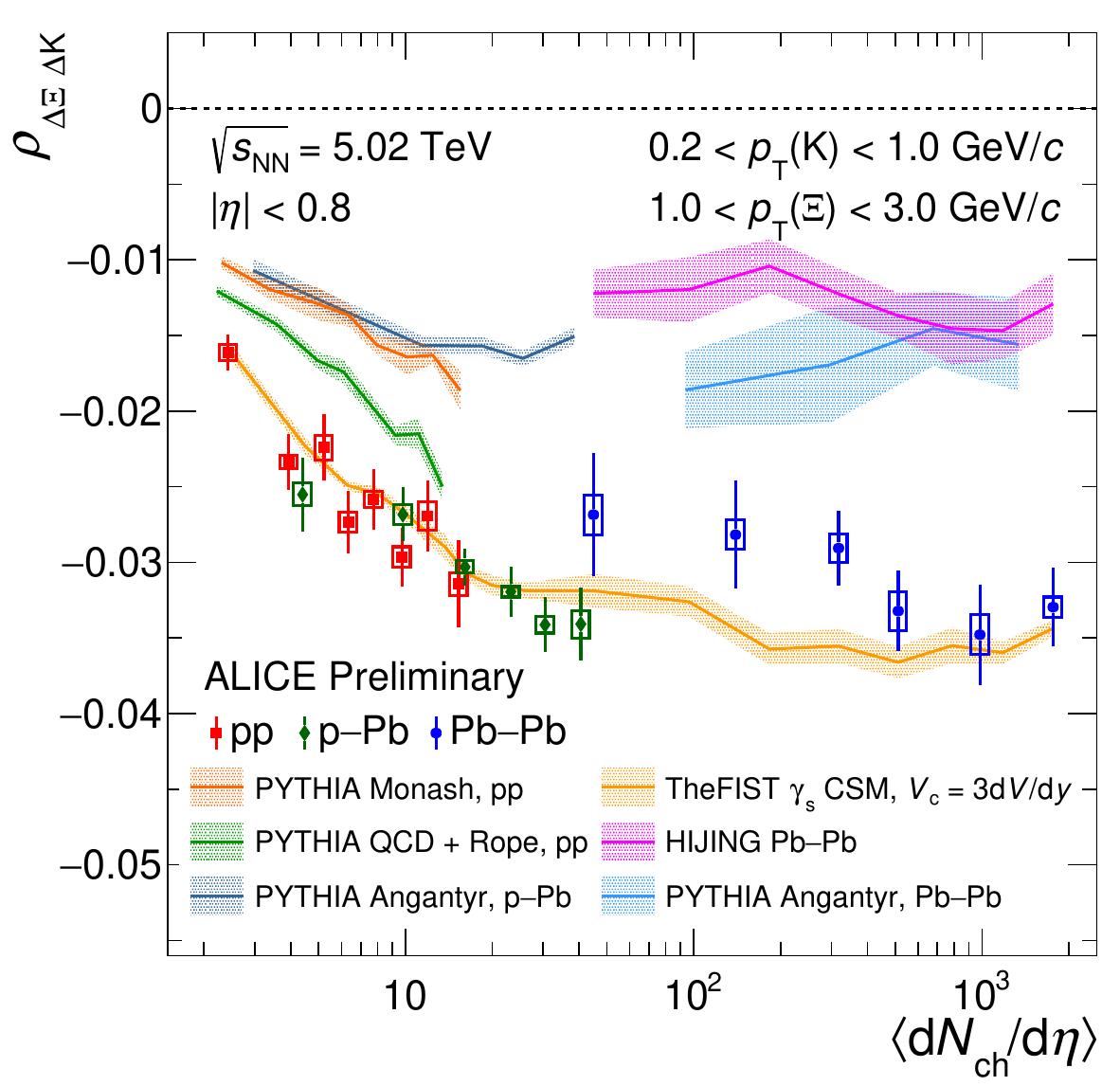}
\caption{Left: ratio between the second- and first-order cumulants of the net-$\Xi^\pm$ number. Right: Pearson correlation between the net-$\Xi^\pm$ and net-kaon number~\cite{PublicNote}. Both observables are reported in pp (red), p--Pb (green), and Pb--Pb collisions (blue), as a function of the average multiplicity of charged particles at midrapidity. Error bars and boxes represent statistical and systematic uncertainties, respectively. The colored bands show the results of model calculations~\cite{PublicNote}.}
\label{fig-netxi}       
\end{figure}
\section{Net-$\Xi^\pm$ and net-kaon number fluctuations}
The canonical SHM model and the string fragmentation mechanisms provide different treatments of charge conservation. In the former, charges are conserved across a finite volume, denoted as correlation volume, $V_\mathrm{c}$. This entails symmetric Pearson correlations of the number of hadrons carrying like-signed and unlike-signed charges. On the other hand, the breaking of the color string connecting a quark to an antiquark implies charge conservation at a local level via the formation of a new quark-antiquark pair. Consequently, the correlation of like-signed hadron numbers is more suppressed than the correlations of unlike-signed hadron numbers.

By measuring the correlation of the number of hadrons of different species, it is possible to observe both the unlike- and like-sign correlations. In the strangeness sector, this study is performed using charged kaons and $\Xi^\pm$ baryons, as they are affected negligibly by autocorrelation effects due to decays of heavy states~\cite{PublicNote}. The results presented in this contribution are extracted from samples of about 900 million pp, 600 million p--Pb, and 400 million Pb--Pb collisions at a center-of-mass energy per nucleon pair $\sqrt{s_\mathrm{NN}}=5.02\ \mathrm{TeV}$ collected by the ALICE Collaboration at the LHC. Charged kaons are identified in the ALICE apparatus using information from the Time Projection Chamber (TPC) and the Time-Of-Flight (TOF) detectors, while $\Xi^\pm$ candidates are reconstructed via the cascade decay channel $\Xi^-\to\Lambda(\to \mathrm{p} + \pi^-) + \pi^-$, and charge conjugate for the $\Xi^+$, and selected via boosted decition trees (BDT). The event-by-event observables are extracted for the net-number of particles, i.e., the difference between the number of particles and that of antiparticles, which are robust against volume fluctuations thanks to the antimatter-matter balance observed at the LHC ~\cite{Rustamov17, MuB}.

The results for the second-to-first-order cumulant ratio of net-$\Xi^\pm$ number and net-$\Xi^\pm$--net-kaon correlation are shown in Fig.~\ref{fig-netxi}. The experimental data show a continuous evolution from small to large collision systems. These results are compared with the aforementioned models, along with HIJING~\cite{Hijing} and Pythia Angantyr~\cite{Angantyr}. All of the models qualitatively predict that the cumulant ratio of net-$\Xi^\pm$ is smaller than the Poissonian baseline of the grand canonical ensemble, corresponding to unity, and that a significant anticorrelation exists between the net-$\Xi^\pm$ and the net-kaon numbers. In both cases, the canonical SHM model implemented in the Thermal-FIST code~\cite{ThermalFist} (orange band), including a strangeness saturation parameter $\gamma_s$, simultaneously describes both of the observables across the analyzed colliding systems using a correlation volume for strangeness and baryon conservation $V_\mathrm{c}=3\mathrm{d}V/\mathrm{d}y$. On the other hand, the models based on the string fragmentation mechanism underpredict both the cumulant ratio of net-$\Xi^\pm$ and the net-$\Xi^\pm$--net-kaon anticorrelation. The former variable is mostly sensitive to unlike-sign strangeness correlations, while the latter is sensitive both to the like- and unlike-sign correlations. Consequently, the model-to-data comparison suggests that in hadronic and heavy-ion collisions strangeness conservation is regulated by long-range rapidity correlations, both for like- and unlike-sign strangeness combinations. Presently, these properties cannot be simultaneously described by models describing strangeness hadronization via color string breaking.

The correlation volume for strangeness and baryon-number conservation is extracted via a combined fit of the net-$\Xi^\pm$ cumulant ratio and the net-$\Xi^\pm$--net-kaon correlation in Pb--Pb with calculations obtained with Thermal-FIST varying $V_\mathrm{c}$. The obtained result is compatible with $3\mathrm{d}V/\mathrm{d}y$, as obtained also in previous studies based on strange-hadron average yields. This result shows that strangeness is conserved over a large volume, similarly to what was observed for the baryon number from the study of net-proton fluctuations~\cite{NetProton}.

\section{Antideuteron--antiproton correlation}
In the previous section, it was shown that event-by-event fluctuations provide a powerful tool to distinguish among different particle-production mechanisms. This strategy can be applied also to the study of (anti)nucleosynthesis~\cite{AntidFluct}. Using antiparticles, it is possible to suppress the contribution of antinuclei produced by the interaction of primary particles with the ALICE apparatus material. To test the applicability of the coalescence model, the antideuteron number is also correlated with the antiproton number measured at half of the transverse momentum, $p_{\mathrm{T}}$, of the antideuteron.

The analysis is based on a sample of about 90 million Pb--Pb collisions at $\sqrt{s_{\mathrm{NN}}}=5.02\ \mathrm{TeV}$. Antideuterons and antiprotons are identified using the particle identification (PID) information provided by the specific energy loss, $\mathrm{d}E/\mathrm{d}x$, in the TPC and particle velocity, $\beta$ measured with the TOF. In this case, volume fluctuation effects are cancelled out by means of centrality bin-width corrections~\cite{VolumeFluctuations}.

\begin{figure}
\centering
\sidecaption
\includegraphics[width=0.48\textwidth]{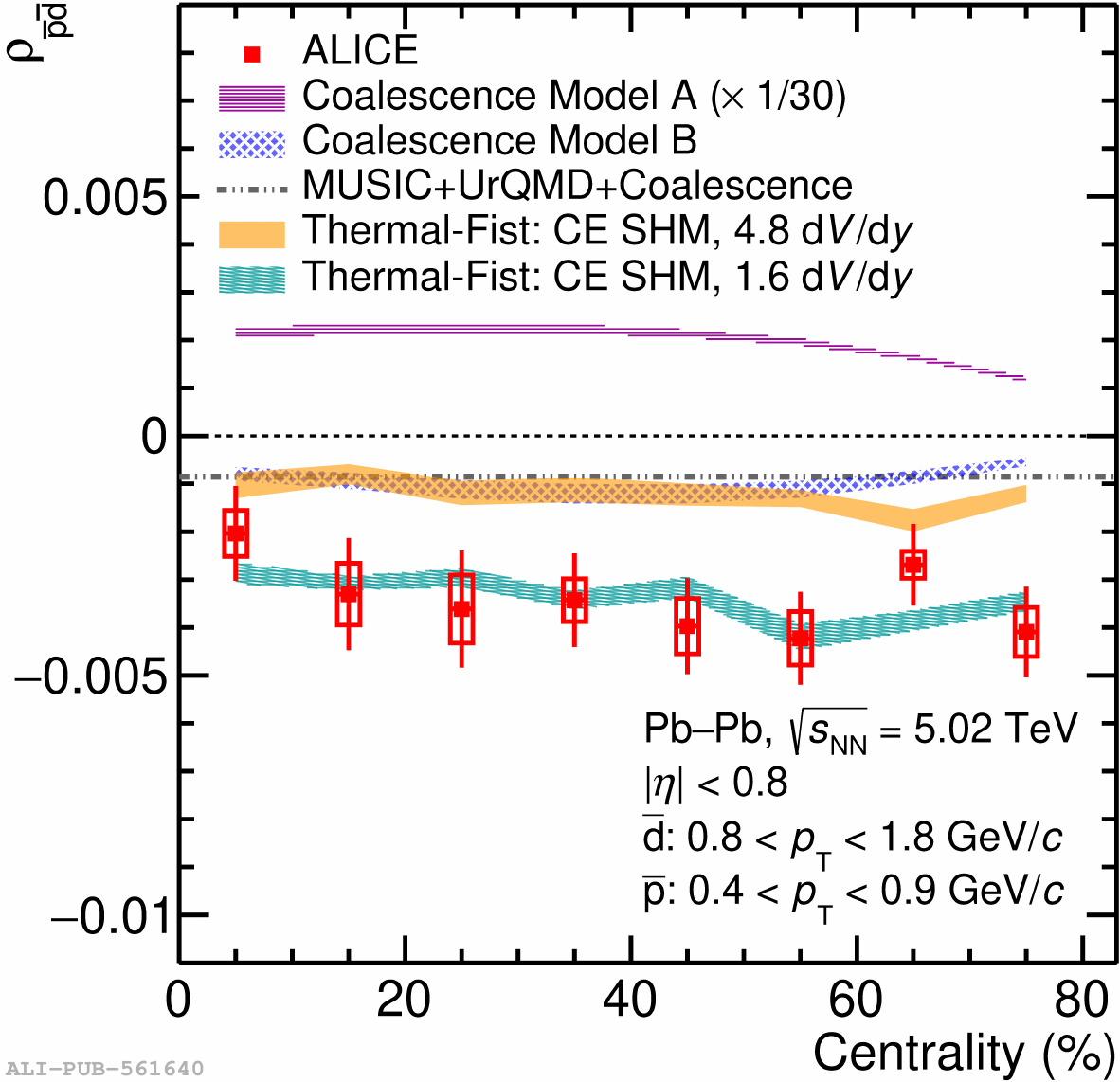}
\caption{Antideuteron-antiproton correlation in Pb--Pb collisions at $\sqrt{s_{\mathrm{NN}}}=5.02\ \mathrm{TeV}$ as a function of the collision centrality~\cite{AntidFluct}. Error bars and boxes depict statistical and systematic uncertainties. The model predictions are shown as bands.}
\label{fig-antid}       
\end{figure}

The results of the antideuteron-antiproton number correlation are shown in Fig.~\ref{fig-antid}. The data show a significant anticorrelation between antideuterons and antiprotons, indicating that baryon-number conservation effects regulate the formation of antideuterons. The data are also compared with calculations of both the canonical SHM model implemented in Thermal-FIST and of simple coalescence models based either on correlated (model A) or independent (model B) production of nucleons~\cite{SimpleCoal}. In addition, improved coalsecence calculations based on the MUSIC+UrQMD predictions~\cite{UrQMDCoal} are shown in grey. The coalsecence model A is ruled out by the measurement, as it predicts a positive correlation, while all other models predict a significant negative correlation. The SHM approach quantitatively describes the data using $V_{\mathrm{c}}=1.6\mathrm{d}V/\mathrm{d}y$ for baryon-number conservation. This volume is smaller than the one extracted from net-proton fluctuations, possibly suggesting that the formation mechanism of antinuclei might differ from the production of other light-flavoured hadrons.

\section{Conclusions}
Event-by-event fluctuations provide new insights in the strangeness hadronization, allowing us to discriminate between models based on either statistical hadronization or string fragmentation, and (anti)nucleosynthesis, where a smaller correlation length for baryon number conservation is detected compared to other light hadrons. These techniques will be further explored with the large data samples collected in the present and upcoming runs of the LHC.

%
%
%

\end{document}